\documentstyle[twocolumn,floats,epsfig,aps,prl]{revtex}

\begin{document}
\widetext
\draft
\twocolumn[\hsize\textwidth\columnwidth\hsize\csname
@twocolumnfalse\endcsname
\title{Cyclic Spin Exchange in Cuprate Spin Ladders}
\author{Tamara S. Nunner$^1$, Philipp Brune$^2$, Thilo Kopp$^1$, Marco Windt$^3$, and Markus Gr\"{u}ninger$^3$}
\address{$^1$EP VI, $^2$TP III, Universit\"{a}t Augsburg, 86135
Augsburg, Germany}
\address{$^3$ II. Physikalisches Institut, Universit\"{a}t zu K\"{o}ln,
50937 K\"{o}ln, Germany}
\date{\today}
\maketitle
\begin{abstract}
We investigate the influence of a cyclic spin exchange $J_{cyc}$ on the one- and
two-triplet excitations of an undoped two-leg $S$=1/2 ladder, using the density
matrix renormalization group (DMRG).
The dispersion of the $S$=0 two-triplet bound state is dramatically reduced by $J_{cyc}$
due to a repulsion between triplets on neighboring rungs.
In (La,Ca)$_{14}$Cu$_{24}$O$_{41}$ a consistent description of both the spin gap and the
$S$=0 bound state requires
$J_{cyc}/J_\perp \! \approx \! 0.20$--0.27 and
$J/J_{\perp} \! \approx \! 1.25$--1.35.
With these coupling ratios the recently developed dynamical DMRG yields an excellent description
of the entire $S$=0 excitation spectrum observed in the optical conductivity,
including the continuum contribution.
\end{abstract}
]
\narrowtext

The antiferromagnetic parent compounds of the high-$T_c$ cuprates are thought to be
the best representatives of the two-dimensional $S$=1/2 square-lattice Heisenberg model.
Understanding their magnetic properties is of utmost importance due to the intimate
relationship of magnetic correlations and high-$T_c$ superconductivity.
Recently, the question how to set up a minimal model which accounts
for these magnetic properties has been readdressed.
High-resolution inelastic neutron scattering experiments performed on the two-dimensional
$S$=1/2 anti\-ferromagnet
${\rm La_2CuO_4}$~\cite{ColdeaHaydenAeppli} exhibit a magnon dispersion at the zone
boundary which cannot be obtained within a nearest-neighbor Heisenberg model.
It has been argued that the inclusion of a cyclic spin exchange term of about
20\% would reproduce this dispersion~\cite{KataninKampf}.
This cyclic spin exchange emerges as a correction to the nearest-neighbor Heisenberg
Hamiltonian in order $t^4/U^3$ from a $t/U$-expansion of the one-band
Hubbard model~\cite{MacDonaldGirvin}.
It is expected to be the dominant correction within a  more realistic three-band
description of the ${\rm CuO}_2$-planes because there
the cyclic permutation of 4 spins on a plaquette can
take place without double occupancy~\cite{MuellerHartmann,SchmidtKuramoto}.
Similar cyclic spin exchange processes have proven to be significant in other systems,
e.g. in cuprate spin chains a ferromagnetic 2-spin cyclic exchange is responsible for
the unusually strong exchange anisotropy~\cite{Kataev}.

Cuprate spin ladders offer an alternative approach to decide about the existence and
potential implications of a cyclic spin exchange term. They are composed of the same
corner-sharing Cu-O plaquettes as the 2D cuprates, thus similar exchange couplings
are expected for all spin products. In fact the inclusion of a cyclic spin exchange
has been suggested in order to explain the smallness of the ladder spin gap observed in
(La,Ca)$_{14}$Cu$_{24}$O$_{41}$~\cite{Brehmer,Matsuda,Johnston}.
However, it is impossible to extract a unique set of coupling parameters or even to
settle the existence of a cyclic spin exchange term from an analysis of the spin
gap only. Here, the optical conductivity $\sigma(\omega)$ \cite{PRL2001,SNS2001} can
provide the missing information. Magnetic excitations can be observed in $\sigma(\omega)$
via the simultaneous excitation of a
phonon~\cite{LorenzanaSawatzky,JureckaBrenig}.
In (La,Ca)$_{14}$Cu$_{24}$O$_{41}$, two peaks in $\sigma(\omega)$ were identified as
$S$=0 bound states of two triplets \cite{PRL2001}.

Here we show that a consistent evaluation of $\sigma(\omega)$ is achieved by inclusion
of $J_{cyc}$. The dispersion of the $S$=0 bound state is very sensitive to the magnitude
of $J_{cyc}$. The combined analysis of both the $S$=0 bound state and the spin gap
establishes a unique set of exchange couplings, namely
$J_{cyc}/J_\perp \! \approx \! 0.20$-0.27,
$J/J_{\perp} \! \approx \! 1.25$-1.35 and $J_\perp \! =\! 950$-1100 cm$^{-1}$ for
(La,Ca)$_{14}$Cu$_{24}$O$_{41}$. For these values, we calculate the entire spectrum
of $\sigma(\omega)$ using the dynamical DMRG~\cite{KuehnerWhite}
and find excellent agreement with new experimental data which yield an improved
estimate of the continuum contribution. This clearly confirms the above value for $J_{cyc}$.
One might speculate that such a sizeable cyclic spin exchange term will have significant
consequences for superconductivity in the doped spin ladders \cite{Schoen,Uehara} since
magnetic and pairing correlations are considered as closely related phenomena \cite{DagottoRice}.

We study an isolated two-leg $S$=1/2 ladder characterized by an antiferromagnetic
Heisenberg Hamiltonian with an additional cyclic spin exchange term:
\begin{eqnarray}
\label{eq:Hamiltonian}
&H& = J_\perp \sum_i {\bf S}_{i,l} {\bf S}_{i,r}
    + J \sum_i \left (
         {\bf S}_{i,l} {\bf S}_{i+1,l} + {\bf S}_{i,r} {\bf S}_{i+1,r}
             \right ) \\
  &+& \! J_{cyc} \! \sum_i \! \frac{1}{4} \! \left( \!
              P^{ }_{(i,l),(i,r),(i+\!1,r),(i+\!1,l)}
          \!\! + \!\! P^{-1}_{(i,l)(i,r)(i+\!1,r)(i+\!1,l)}
                           \right )  \nonumber
\end{eqnarray}
where $J_\perp$ and $J$ denote the rung and leg couplings, the
index $i$ refers to the rungs, and $l$, $r$ label the two legs.
The cyclic permutation operator $P_{1234}$  for 4~spins on a plaquette
is given by:
\begin{eqnarray}
 &P&^{ }_{1234} + P^{-1}_{1234} =
\\
& & 4 ({\bf S}_1 {\bf S}_2)({\bf S}_3 {\bf S}_4)
       + 4 ({\bf S}_1 {\bf S}_4)({\bf S}_2 {\bf S}_3)
       - 4 ({\bf S}_1 {\bf S}_3)({\bf S}_2 {\bf S}_4)
   \nonumber \\
       & &
       + {\bf S}_1 {\bf S}_2 + {\bf S}_3 {\bf S}_4
       + {\bf S}_1 {\bf S}_4 + {\bf S}_2 {\bf S}_3
       + {\bf S}_1 {\bf S}_3 + {\bf S}_2 {\bf S}_4
                     \, . \nonumber
\end{eqnarray}

The influence of $J_{cyc}$ can easily be understood in terms of the
representation of the spin operators by rung singlets $s_i^\dag$ and
triplets $t_{i,\alpha}^\dag$ ($\alpha=x, y, z$)
\cite{SachdevBhattGopalanRice,ConstraintCyclic}:
\begin{eqnarray}
\label{eq:HamiltonianTriplet}
H   = \sum_i & & \left \{
     {\tiny \frac{9}{16} J_{cyc}}
    - \frac{3}{4} (J_\perp + \frac{1}{2} J_{cyc}) s_i^\dag s_i
\right.  \\
 &  &+ \frac{1}{4}(J_\perp -\frac{11}{2} J_{cyc})
                   \sum\nolimits_\alpha t_{i,\alpha}^\dag t_{i,\alpha}
\nonumber \\
 & &+ \frac{1}{2} (J + \frac{1}{2} J_{cyc})
                  \sum\nolimits_\alpha \left [
                       s_i^\dag s_{i+1} t_{i+1,\alpha}^\dag t_{i,\alpha}
                       + {\rm H.c.} \right ]
\nonumber \\
& &   + \frac{1}{2} (J - \frac{1}{2} J_{cyc})
                  \sum\nolimits_\alpha \left [
                       s_i^\dag s_{i+1}^\dag t_{i+1,\alpha} t_{i,\alpha}
                       + {\rm H.c.} \right ]
\nonumber \\
 & & + \frac{1}{4} (J + \frac{1}{2} J_{cyc})
                  \sum\nolimits_{\alpha,\beta} (1-\delta_{\alpha
                                                 \beta}) \times
\nonumber \\
& & \!\!\!\!\!\!\!\! \left [ t_{i,\alpha}^\dag t_{i+1,\beta}^\dag t_{i+1,\alpha}
      t_{i,\beta}
       - t_{i,\alpha}^\dag t_{i+1,\alpha}^\dag t_{i+1,\beta} t_{i,\beta}
                       + {\rm H.c.} \right ]
\nonumber \\
 & & +  \left. J_{cyc} \sum\nolimits_{\alpha,\beta}
             t_{i,\alpha}^\dag t_{i,\alpha} t_{i+1,\beta}^\dag t_{i+1,\beta}
       \right \} \, . \nonumber
\end{eqnarray}

\begin{figure}[t!]
\begin{center}
\epsfig{figure=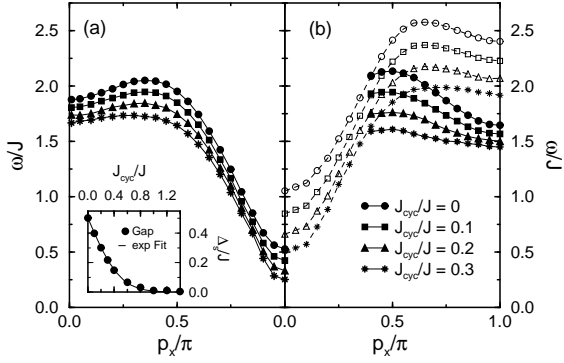,width=7.5cm}
\end{center}
\caption{
DMRG results for a N=80-site ladder with $J\!=\!J_\perp$ and $0\leq J_{cyc}/J \leq 0.3$.
(a) One-triplet dispersion. The inset shows the spin gap $\Delta_s$ (extrapolated to
$N\!=\!\infty$ {\protect \cite{FiniteSizeScaling}}) as a function of
$J_{cyc}/J$ together with an exponential fit.
(b) Lower edge of the two-triplet continuum (open symbols)
and the $S=0$ two-triplet bound state (full symbols).}
\label{fig:DispersionCyclic}
\end{figure}

The consideration of $J_{cyc}$ primarily renormalizes the coupling strengths
of the individual terms in the original Heisenberg Hamiltonian. The only new contribution
is the last term, a repulsive interaction between triplets on neighboring rungs.
With respect to the renormalization of the coupling constants the strongest effect is a
reduction of the triplet on-site energy. Consequently the spin gap as well as the energy
of the entire one-triplet dispersion are reduced.
In Fig.~\ref{fig:DispersionCyclic}(a) we show the one-triplet dispersion, obtained
by the Lanczos-vector method~\cite{Hallberg,LanczosDetails} for a N=80-site ladder
with $J\!=\!J_\perp$ and different values of $J_{cyc}$. The shape of the one-triplet
dispersion remains qualitatively unchanged. The strongest suppression affects
the minimum ($p_x \! = \! \pi$) and the maximum (near $p_x \! \approx \! \pi/3$).
Nevertheless, there still is a local minimum at $p_x$=0
even for $J_{cyc}/J \! = \! 0.3$, in contrast
to the exact diagonalization results for a N=24-site ladder in Ref.~\cite{Brehmer}.
Extrapolating to an infinite ladder~\cite{FiniteSizeScaling}, we find that
the spin gap $\Delta_s=E_0(N\!=\!\infty,S\!=\!1)-E_0(N\!=\!\infty,S\!=\!0)$
decreases exponentially (inset of Fig.~\ref{fig:DispersionCyclic}(a)),
but we cannot conclusively decide from our present data, whether the spin gap indeed
vanishes at $J_{cyc}/J \approx 1.2$, as has been proposed in Ref.\cite{HondaHoriguchi}.

Similar considerations also apply to the two-triplet excitations. With increasing
$J_{cyc}$ the lower edge of the two-triplet continuum (open symbols in
Fig.~\ref{fig:DispersionCyclic}(b)) shifts corresponding to the one-triplet dispersion.
In addition, however, a repulsive interaction between neighboring triplets is introduced
by $J_{cyc}$, as mentioned above (see Eq.~(\ref{eq:HamiltonianTriplet})). This reduces
the binding energy of the $S$=0 two-triplet bound state. As a consequence the width of
the bound-state dispersion decreases dramatically (full symbols in
Fig.~\ref{fig:DispersionCyclic}(b)). Evidently the dispersion of
the $S$=0 two-triplet bound state is very sensitive to the magnitude of $J_{cyc}$ and thus
provides a suitable quantity to probe the existence and magnitude of a cyclic spin exchange.

\begin{figure}[t!]
\begin{center} 
\epsfig{figure=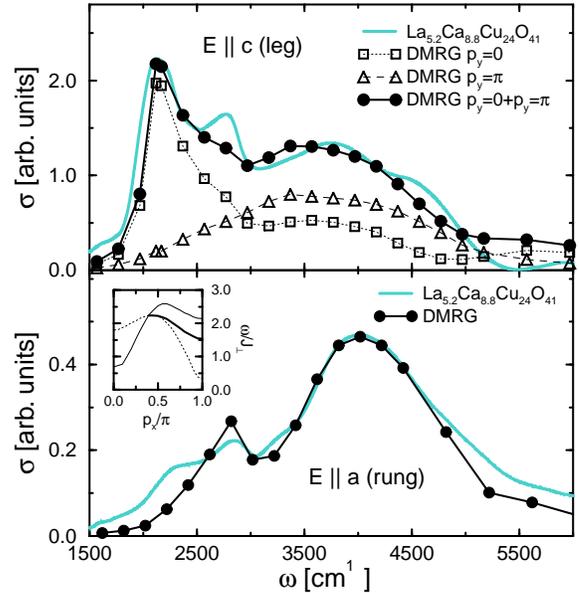,width=7.5cm}
\end{center}
\caption{
Symbols: Correction-vector results for the optical conductivity $\sigma(\omega)$ of
an N=80-site ladder with polarization of the electrical field E along the legs (top panel)
and along the rungs (bottom panel), using $J/J_\perp$=1.3, $J_{cyc}/J_\perp$=0.2,
$J_\perp$=1000 cm$^{-1}$, $\omega_{ph}^{\rm leg}$=570 cm$^{-1}$,
$\omega_{ph}^{\rm rung}$=620 cm$^{-1}$ {\protect \cite{phonon}}
and a finite broadening of $\delta=0.1 J_\perp$.
The leg polarization contains two contributions, where the two legs are in-phase ($p_y$=0)
or out-of-phase ($p_y\!=\!\pi$) with each other.
Gray lines: $\sigma(\omega)$ of
${\rm La_{5.2}Ca_{8.8}Cu_{24}O_{41}}$ at T=4~K {\protect \cite{exp}}.
Inset: one-triplet dispersion (dashed line),
lower edge of the two-triplet continuum (thin solid line) and $S$=0 bound state
(thick solid line) for the above parameters.
}
\label{fig:OptLeitLegRung}
\end{figure}

In (La,Ca)$_{14}$Cu$_{24}$O$_{41}$ the $S$=0 bound state has been observed
in the optical conductivity $\sigma(\omega)$ via phonon-assisted two-triplet
absorption \cite{PRL2001}. Although the total momentum of the excitation has to be zero,
the simultaneous excitation of a phonon with arbitrary momentum
permits to measure a weighted average of the $S$=0 excitations from the entire Brillouin zone.
In Fig.\ \ref{fig:OptLeitLegRung} we show the magnetic contribution to $\sigma(\omega)$
in La$_{5.2}$Ca$_{8.8}$Cu$_{24}$O$_{41}$ \cite{exp}.
For polarization of the electrical field parallel to the legs, the
two peaks in $\sigma_{\rm leg}(\omega)$ at $\omega_1$=2140 cm$^{-1}$ and
$\omega_2$=2780 cm$^{-1}$
correspond to van Hove singularities resulting from the maximum $\omega_{\rm max}$ and
minimum $\omega_{\rm min}$ of the $S$=0 bound-state dispersion at $p_x \! \approx \! \pi/2$
and $p_x \! = \! \pi$, respectively
(see Fig.~\ref{fig:DispersionCyclic}(b)).
The broad features at higher energies represent the continuum contribution.

\begin{figure}[b!]
\begin{center}
\epsfig{figure=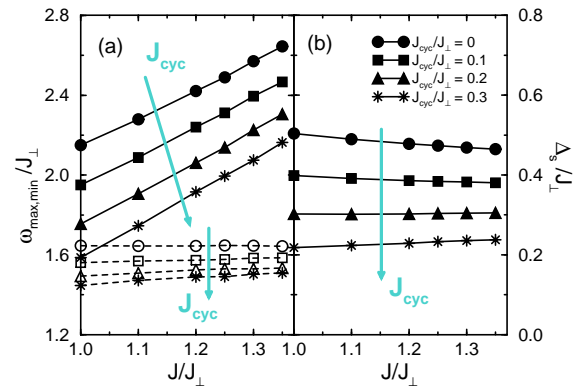,width=7.5cm, clip=}
\end{center}
\caption{
(a) Maximum of the $S$=0 bound-state dispersion at $p_x \approx \pi/2$
(full symbols), minimum at $p_x=\pi$ (open symbols)
and (b) spin gap $\Delta_s$ as a function of $J/J_\perp$ for different values of
$J_{cyc}/J_\perp$.
All results have been extrapolated to N=$\infty$ {\protect \cite{FiniteSizeScaling}}.
}
\label{fig:PeakPosGap}
\end{figure}

In order to determine the exchange couplings for La$_{5.2}$Ca$_{8.8}$Cu$_{24}$O$_{41}$
we have calculated
$\omega_{\rm max}$ and $\omega_{\rm min}$
for different coupling ratios $J/J_\perp$ and $J_{cyc}/J_\perp$ using the
Lanczos-vector method~\cite{LanczosDetails}. By extrapolation to an infinite
ladder~\cite{FiniteSizeScaling} we obtain the values displayed in
Fig.~\ref{fig:PeakPosGap}(a). The corresponding results for the spin gap $\Delta_s$
are shown in Fig.~\ref{fig:PeakPosGap}(b). In the range depicted in
Fig.~\ref{fig:PeakPosGap} these values may be approximated within 5\% by
$\omega_{\rm min} \! \approx \! 1.64 J_\perp - 0.54 J_{cyc}$,
$\omega_{\rm max} \! \approx \! 0.61 J_\perp - 1.87 J_{cyc} + 1.53 J$ and
$\Delta_s \! \approx \! 0.48 J_\perp - 0.84 J_{cyc}$.
In principle the three experimental quantities $\omega_1$, $\omega_2$ and
$\Delta_s \! \approx \! 280$ cm$^{-1}$~\cite{Matsuda} should suffice to determine
$J_{cyc}$, $J$ and $J_\perp$. However, $\omega_1$ and $\omega_2$ depend also
on the frequency $\omega_{ph}$ of the phonon involved in the optical absorption
process, i.e.,
$\omega_1 \!=\! \omega_{\rm min} + \omega_{ph}^{p_x=\pi}$ and
$\omega_2 \!=\! \omega_{\rm max} + \omega_{ph}^{p_x=\pi/2}$.
Making use of the relative insensitivity of $\Delta_s$ and $\omega_{\rm min}$
on $J/J_\perp$, we can determine the magnitude of the cyclic spin exchange
without consideration of $\omega_{2}$ and hence omit any effect of a possible
phonon dispersion.
For a conservative estimate of the Cu-O bond-stretching phonon frequency
$\omega_{ph}^{p_x=\pi} \! = \! 600 \pm 100$cm$^{-1}$ we obtain
$J_{cyc}/J_\perp \! = \! 0.20-0.27$
with $J_\perp \approx 950 - 1100$cm$^{-1}$.
The dispersion of the phonon enters only in the determination of the
coupling ratio $J/J_\perp$. Allowing for a dispersion of
$\omega_{ph}^{p_x=\pi/2} \! -  \omega_{ph}^{p_x=\pi} \approx \pm 50$ cm$^{-1}$
we obtain $J/J_\perp \approx 1.25-1.35$.

Up to now we have derived a set of exchange couplings which consistently describes
the spin gap and the $S$=0 two-triplet bound state. Now we compare the spectral
density calculated for this parameter set with the line shape of the entire spectrum
observed in $\sigma(\omega)$. This provides a thorough test whether we indeed have
found the minimal model which captures all relevant properties. We consider
an isolated ${\rm Cu_2O_3}$-ladder, for which the major contribution to $\sigma(\omega)$
results from the
simultaneous excitation of two neighboring spin-flips plus
a Cu-O bond-stretching phonon:
\begin{equation}
\sigma( \omega) \sim  - \omega \sum_{p_x} \sum_{p_y=0,\pi}
          f_p \, {\rm Im} \langle \langle \delta B_{-p} ; \delta B_p
                  \rangle \rangle_{( \omega- \omega_{ph})}
\label{eq:sigma}
\end{equation}
where
$\delta B_{p}^{\rm leg} \!\! = \!\! \frac{1}{N} \! \sum_i \! \sum_{k=l\!,r} e^{ipr_{i\!,k}} \! \left (
      {\bf S}_{i,k} {\bf S}_{i+1\!,k}
        \! - \! \langle {\bf S}_{i,k} {\bf S}_{i+1\!,k} \rangle  \right )$
and
$\delta B_{p}^{\rm rung} \!=\! \frac{1}{N} \sum_i e^{ipr_i} \left (
      {\bf S}_{i,l} {\bf S}_{i,r}
        - \langle {\bf S}_{i,l} {\bf S}_{i,r} \rangle \right )$
are the spin-flip operators for polarization of the electrical field along
the legs and the rungs, respectively.
To lowest order (4th order in the Cu-O hopping $t_{pd}$), the dominant contribution to
the phonon form factor $f_p^{\rm leg}$ comes from the in-phase and the out-of-phase
stretching modes of the O-ions on the legs, whereas for $f_p^{\rm rung}$ the out-of-phase
stretching mode and the vibration of the O-ion on the rung are taken into account:
\begin{equation}
\label{eq:formfactor}
f_p^{\rm leg} = 8 \sin^4 (\frac{p_x}{2}) \,\,\,\, ,\,\,\,\,
f_p^{\rm rung} = 8 \sin^2 (\frac{p_x}{2}) + 4
\end{equation}

So far we have used the Lanczos-vector method (LVM)~\cite{Hallberg} to
calculate the dispersion of the bound state. This was justified by the
fact that the LVM uses the first
Lanczos vectors as target states besides
the ground state and consequently represents the bound state quite accurately.
However, in order to calculate the high-frequency range of $\sigma(\omega)$ the LVM
is not reliable. We use the correction-vector method~\cite{KuehnerWhite,DMRGDetails},
because here the density matrix is optimized to represent the correction vector for
each frequency point separately.

In Fig.~\ref{fig:OptLeitLegRung} we compare $\sigma(\omega)$ for polarization along the legs
and along the rungs, obtained by the correction vector method for a ladder with N=80-sites,
with the experimental result of ${\rm La_{5.2}Ca_{8.8}Cu_{24}O_{41}}$
\cite{exp}.
Here we have used $J/J_\perp$=1.3 and $J_{cyc}/J_\perp$=0.2 with $J_\perp=1000\,\,{\rm cm}^{-1}$
\cite{phonon}.
For polarization along the legs, $\sigma(\omega)$ contains two contributions, where
the two legs are excited in-phase ($p_y=0$) or out-of-phase
($p_y=\pi$).
The in-phase mode contains two- and four-triplet excitations, as can be
inferred from the rung-triplet representation of the spin-flip operator
${\bf S}_{i,l}{\bf S}_{i+1,l}+{\bf S}_{i,r}{\bf S}_{i+1,r}$ in
Eq.~(\ref{eq:HamiltonianTriplet}).
It includes the $S$=0 two-triplet bound state as discussed above.
The out-of-phase mode reflects excitations of three triplets due to
${\bf S}_{i,l}{\bf S}_{i+1,l}-{\bf S}_{i,r}{\bf S}_{i+1,r}=
\frac{i}{2} \sum_{\alpha,\beta,\gamma}
\! \epsilon_{\alpha \beta \gamma} \! \left \{
t_{i,\alpha}^\dag t_{i+1,\beta}^\dag t_{i+1,\gamma}^{} s_i
\!+  t_{i,\beta}^\dag t_{i+1,\alpha}^\dag s_{i+1} t_{i,\gamma}
\!\!-\! {\rm H.c.} \right \}$
and hence contributes only to the high-energy continuum
 excitations~\cite{tobepublished}.
For polarization along the rungs, $\sigma(\omega)$ contains only
two-triplet excitations (see Eq.~(\ref{eq:HamiltonianTriplet}) for the
rung-triplet representation of ${\bf S}_{i,l}{\bf S}_{i,r}$)
and the lower bound state is suppressed due to a selection  rule~\cite{PRL2001}.

Comparing the DMRG results for polarization along the legs with the experimental
spectra we find excellent agreement within the range of the bound state (which has
been used in order to fix the coupling constants), and also
with respect to the overall line shape of the high-energy continuum excitations.
For this polarization the experimental determination of the precise spectral weight
of the continuum is difficult due to an electronic background \cite{PRL2001,SNS2001,exp}.
The small differences between experiment and theory in the continuum range are certainly
within the experimental error bars.
For polarization along the rungs, where experimentally the magnetic contribution to $\sigma(\omega)$
can be identified unambiguously and with high precision \cite{PRL2001,SNS2001}, the continuum is
reproduced nearly perfectly.
This provides strong evidence that a nearest-neighbor Heisenberg Hamiltonian with
$J/J_\perp$=1.3 and an additional cyclic spin exchange of $J_{cyc}/J_\perp$=0.2
represents the magnetic excitations of the spin-ladder compound
(La,Ca)$_{14}$Cu$_{24}$O$_{41}$.

In spite of the overall impressive consistency of experiment and theory, some
discrepancies are visible:
(i) In $\sigma_{\rm leg}(\omega)$ the upper bound state (for $p_x \approx \pi/2$) is suppressed
much stronger in the DMRG calculation than in the experimental
spectrum.
(ii) In $\sigma_{\rm rung}(\omega)$ the experimental data displays a pronounced
shoulder at
about 2200 cm$^{-1}$, i.e., slightly above the suppressed lower bound state, which is
not present in our DMRG calculation.
We propose that these discrepancies are related to different phonon form factors
which arise due to the actual arrangement of the ladders within
the trellis lattice,
and in higher order (with respect to Cu-O hopping) in the external
photon-electron-phonon vertex.
Considering, e.g., an additional term $\propto \sin^2(p_x/2)$ in
$f_p^{\rm leg}$ will enhance
the weight of the upper bound state in $\sigma_{\rm leg}(\omega)$ with
respect to the lower one.
Such considerations, however, depend on details of the non-local excitation process in these ladder
systems and are therefore not of fundamental relevance.

In summary, we have demonstrated that the inclusion of a cyclic spin exchange $J_{cyc}$ is
necessary to obtain a consistent interpretation of optical experiments and the magnitude of
the spin gap in (La,Ca)$_{14}$Cu$_{24}$O$_{41}$.
The dispersion of the $S$=0 bound state is very sensitive to the magnitude of $J_{cyc}$.
Quantitative investigation yields
$J_{cyc}/J_\perp \! \approx \! 0.20$-0.27 and $J/J_\perp \! \approx \! 1.25$-1.35
with $J_\perp \! \approx \! 950$-1100 cm$^{-1}$.
With these parameters we obtain excellent agreement with the experimental spectra of
${\rm La_{5.2}Ca_{8.8}Cu_{24}O_{41}}$, both for the bound state and the continuum.
This gives strong evidence that we indeed have identified the minimal model which
captures all relevant magnetic properties.

We would like to thank  M.~Greiter, A.~P.~Kampf and G.~S.~Uhrig
for stimulating discussions.
This project was supported by the DFG (SFB 484 and SFB 608)
and by the BMBF (13N6918/1).


\begin{references}
\bibitem{ConstraintCyclic}
In the transformation of the cyclic term explicit use was made of the constraint:
$s_i^\dag s_i + \sum_\alpha t_{i,\alpha}^\dag t_{i,\alpha} = 1$.
\bibitem{LanczosDetails}
All Lanczos vector calculations were performed with $3-5$ iterations
of the finite-lattice algorithm and a maximum of 200-300 states.
Four Lanczos vectors were used as target states.
All calculations were performed with
open boundary conditions. A Parzen filter was used to
minimize edge effects for the calculation of correlation functions.
\bibitem{FiniteSizeScaling} For all finite size extrapolations a
polynomial in 1/N was used, where the 1/N-term was
excluded. This is the correct finite size scaling for the
$S$=1 chain~\cite{SorensenAffleck}.
\bibitem{DMRGDetails}
Here Ramasesha's algorithm\cite{Ramasesha} has been used instead of the conjugate
gradient method.
\bibitem{phonon}
Experimentally, the frequency of the upper bound state $\omega_2$ is $\approx$ 60 cm$^{-1}$
higher in $\sigma_{\rm rung}(\omega)$ than in $\sigma_{\rm leg}(\omega)$,
thus two different phonon frequencies are used in Fig.~\ref{fig:OptLeitLegRung}.
\end{references}
\end{document}